\documentclass{elsart5p}

\usepackage{graphics}
\usepackage{graphicx}
\usepackage{amssymb}
\begin{document}

\begin{frontmatter}

% Title, authors and addresses

% use the thanksref command within \title, \author or \address for footnotes;
% use the corauthref command within \author for corresponding author footnotes;
% use the ead command for the email address,
% and the form \ead[url] for the home page:
% \title{Title\thanksref{label1}}
% \thanks[label1]{}
% \author{Name\corauthref{cor1}\thanksref{label2}}
% \ead{email address}
% \ead[url]{home page}
% \thanks[label2]{}
% \corauth[cor1]{}
% \address{Address\thanksref{label3}}
% \thanks[label3]{}

\title{Hubbard-Thomas-Fermi Theory of Transition Metal Oxide Heterostructures}
%
% use optional labels to link authors explicitly to addresses:
% \author[label1,label2]{}
% \address[label1]{}
% \address[label2]{}

\author[AA]{Wei-Cheng Lee\corauthref{Lee}},
\ead{leewc@mail.utexas.edu}
\author[AA]{A. H. MacDonald}

\address[AA]{Department of Physics, The University of Texas at Austin, Austin, TX 78712}

\corauth[Lee]{Corresponding author. Tel: (512)471-5426}

\begin{abstract}
We demonstrate that the charge distributions in Hubbard-model representations of 
transition metal oxide heterojucntions can be described by a Thomas-Fermi theory in which 
the energy is approximated as the sum of the electrostatic energy and the uniform three-dimensional Hubbard model energy per site
at the local density equals to a constant. When charged atomic layers in the oxides are approximated as 
two-dimensional sheets with uniform charge density, the electrostatic energy is simply evaluated. 
We find that this Thomas-Fermi theory can reproduce results obtained from full Hartree-Fock theory
for various different heterostructures. We also show explicitly how Thomas-Fermi theory can be used
to estimate some key properties qualitatively.
\end{abstract}


\end{frontmatter}
Recent advances in techniques for layer-by-layer growth of transition metal oxides promise new types of heterostructures 
between strongly-correlated materials
\cite{ohtomo,ohtomo2,nakagawa} and has motivated many theoretical studies devoted
to exploring their physical properties \cite{okamoto, lee, lee2} using simple Hubbard models augmented by a 
long-range Coulombic interaction term.
The charge distribution near a the transition metal oxide heterostructures (TMOHs) is usually studied using Hartree-Fock theory or
dynamical mean-field theory with satisfactory results.  We have found that the same results can be reproduced
much more simply using a Thomas-Fermi theory (TFT) in which all energy contributions other than the electrostatic one 
are approximated by a local density approximation \cite{lee,lee2}. In this paper, 
we derive TFT specifically for the generalized Hubbard models of TMOHs from general consideration and demonstrate how it can be used to estimate some 
key properties of the TMOHs.

The simplest toy model to investigate the TMOHs is: $H=H_U+H_C$, where $H_U$ is the single-band Hubbard model describing 
on-site correlations and $H_C$ is the 
long-ranged Coulomb interaction. Following the spirit of TFT, we can write down TF equation as:
\begin{equation}
\mu(\rho[\vec{r}])+v_H(\vec{r})=const.
\end{equation}
where $\mu(\rho)$ and $v_H(\vec{r})$ are the chemical potential of the Hubbard
model for a uniform system with density $\rho$  
and the electrostatic potential at position $\vec{r}$. 
Since the transition metal oxides have a natural layered structures, we allow the average density to 
vary from layer to layer near the heterojunction.  It is usually a good approximation to 
ignore the periodic but non-uniform density variation within each layer in calculating the electrostatic potential.
In this approxmation both $\mu(\vec{r})$
and $v_H(\vec{r})$ depend only 
on the position along the growth direction $\hat{z}$ and the TF 
equation reduces to:
\begin{equation}
\mu(\rho(z))+v_H(z)=const.
\end{equation}
where $\rho(z)$ is the average electron density at $z$. 

The electrostatic potential $v_H$ can be evaluated from self-consitently determined densities by solving the 
Poisson equation. In the continuum limit each layer can be considered as a two-dimensional sheet with uniform 
effective charge density $\sigma$. Consequently, the electrostatic potential at $z$ is contributed by all charged layers at $z'$ with effective charge density $\sigma(z')$:
\begin{equation}
v_H(z)=\sum_{z'} v_H(z',z)=2\pi U_c\sum_{z'} \sigma(z')\vert z-z'\vert
\end{equation}
where $U_c=e^2/\epsilon a$ is the strength of long-ranged Coulomb interaction, and $a$ is the lattice constant. 
The next thing is to evaluate the chemical potential $\mu(\rho(z))$.
Ideally $\mu(\rho(z))$ should be obtained from exact results for the homogeneous three-dimensional Hubbard model.
Since this is not available some approximations are required. 
The simpliest choice is the Hartree-Fock approximation. Since Hartree-Fock theory can have separate solutions 
for antiferromagnetic (AFM) and ferromagnetic (FM) states(among others),
we can solve TF equation for both cases. It has been shown that TFT accurately reproduces the charge 
distribution along the growth direction obtained by Hartree-Fock theory for modulation-doped Mott-insulator-Mott-insulator and Mott-insulator-band-insulator
heterostructures\cite{lee}. 

\begin{figure}
\begin{center}
\includegraphics{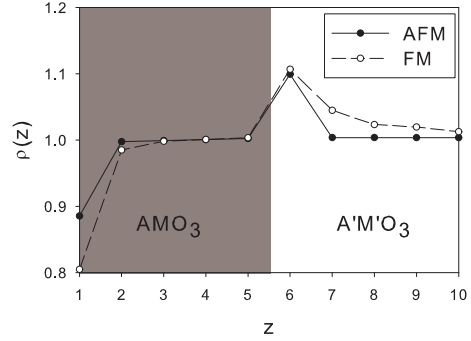}
\end{center}
\caption{\label{fig1} Electron distribution calculated by TFT for AFM and FM states with $U_1/t=U_2/t=20$, $U_c=0.8$, $N_p=5$, and $N=10$ for a polar ($AMO_3$)-nonpolar 
($A'M'O_3$) heterostructure.}
\end{figure}

Recently the authors proposed that electronic interface reconstruction (EIR), 
a charge tranfer from the outmost surface layer to the interface, can occur in a polar 
($AMO_3$)-nonpolar ($A'M'O_3$) peroskite Mott-insulator heterostructure\cite{lee2}. 
The effective charge densities are $\sigma=1$ for the $AO$ layers, $0$ for $A'O$ layers,
$\rho(z)$ for the $MO_2$ layers, and $(\rho(z)-1)$ for $M'O_2$, where $\rho(z)$ is the average
 electron density on the $d$-band of transitional metal ions $M(M')$.
We consider a film with $N_p$ pairs of $AO-MO_2$ layers and $(N-N_p)$ pairs of
$A'O-M'O_2$ layers. The electrostatic potential for this case is\cite{lee2}:
\begin{equation}
\frac{v_H(z)}{2\pi\,U_c}=\sum_{z_A}\vert z-z_A\vert - \sum_{z'\neq z}\left[\rho(z')-\theta(z'-N_p)\right]\vert z'-z\vert
\end{equation}
where $z_A=1.5,2.5,...,N_p+0.5$, $(z,z')=1,2,...,N$, and $\theta(z'-N_p)$ is the Heaviside step function.
The charge distribution calculated from TFT is given in Fig.\ref{fig1}, and is nearly identical to the 
results of Hartree-Fock theory presented 
in the Fig. 2(b) of Ref\cite{lee2}. The success of TFT for different TMOHs is remarkable. The critical reason for this success is that the charge 
distribution in these TMOHs is a consequence of competition between the on-site correlation energy (represented by $\mu(\rho(z))$) and the spatial 
long-ranged Coulomb field (represented by $v_H(z)$), and these two dominant energy scales are correctly captured by TFT. This feature also promises that 
if the chemical potential $\mu(\rho(z))$ can be obtained more realistically by some sophisticated approaches for the homogeneous three-dimensional Hubbard model, 
TFT can also predict the correct charge distribution for most TMOHs without heavy computation.

Another advantage of TFT emerges if we want to estimate some key properties qualitatively.
It is known, for example\cite{lee2}, that a minimum value of 
$N_p$ is required for EIR to occur. This minimum value can be estimated easily using TFT. Since only the 
outer most surface and the interface layer have electron densites different from 1 when EIR first occurs, we consider:
\begin{equation}
\mu_{U_1}(\rho(z_1))+v_H(z_1) = \mu_{U_2}(\rho(z_i))+v_H(z_i)
\label{tf1}
\end{equation}
where $z_1=1$ is the position of outmost layer and $z_i=N_p+1$ is that of doped interfacial layer by our convention. $U_1$ ($U_2$) is the Hubbard $U$ parameter for polar 
(nonpolar) material. We put $\rho(z_1)=1-x$, $\rho(z_i)=1+x$, and $\rho(z)=1$ for all other values of $z$, then we can evaluate $v_H(z_1)$ and $v_H(z_i)$: 
\begin{equation}
\begin{array}{l}
\displaystyle
\frac{v_H(z_1)}{2\pi U_c}=\sum_{z_A}\vert z_1-z_A\vert - \sum_{z'=2}^{N_p}\vert z'-z_1\vert - x\vert z_i-z_1\vert\\[2mm]
\displaystyle
\frac{v_H(z_i)}{2\pi U_c}=\sum_{z_A}\vert z_i-z_A\vert - \sum_{z'=2}^{N_p}\vert z'-z_i\vert - (1-x)\vert z_i-z_1\vert
\end{array}
\label{sd}
\end{equation}
It is straightforward to compute: $\sum_{z_A}\vert z_1-z_A\vert=\sum_{z_A}\vert z_i-z_A\vert=\frac{N_p^2}{2}$, 
$\sum_{z'=2}^{N_p}\vert z'-z_1\vert = \sum_{z'=2}^{N_p}\vert z'-z_i\vert=\frac{N_p^2-N_p}{2}$, and $\vert z_i-z_1\vert = N_p$.
So if we require $x\geq 0$, we find $N_p$ should be: $N_p \geq \frac{\mu_{U_2}(1^+)-\mu_{U_1}(1^-)}{2\pi\,U_c}$.
The insight revealed by this expression has been discussed in Ref\cite{lee2}. 
A similar analyis has been discussed earlier\cite{lee}.

In summary, we have derived a TFT appropriate for Hubbard model descriptions of TMOHs and 
applied it to cubic peroskite heterojunction systems. We have shown that 
this TFT can be used in an appropriate way to estimate some key properties
which are difficult to treat analytically in Hartree-Fock theory and dynamical mean-field theory. 
The applicability of TFT indicates that the on-site correlation and long-ranged Coulomb interaction
are the most dominant energy scales in TMOHs.  For this reason 
it is possible to create new types of two-dimensional systems which
inherit strong-correlation features from the Mott insulator near interface between TMOs.

{\it Acknowledgement} This work is supported in part by the Welch Foundation and by the National Science Foundation 
under grant DMR-0606489.

\end{document}